\documentclass[aps,prd,twocolumn,superscriptaddress,amssymb,eqsecnum,showpacs,showkeyes,secnumarabic,graphics,floatfix,nofootinbib,tightenlines,longbibliography]{revtex4-1}
\usepackage{relsize}
\usepackage{blindtext}
\usepackage{enumitem}
\usepackage{xcolor}
\usepackage{graphicx} 
\usepackage{epsfig}
\usepackage{bm}
\usepackage{cancel}
\usepackage{dcolumn}
\usepackage{amsmath} 
\usepackage{hhline} 
\usepackage[utf8]{inputenc}
\usepackage[breaklinks=true,colorlinks=true,
linkcolor=blue,urlcolor=blue,citecolor=cyan,
bookmarks=true,bookmarksopenlevel=2]{hyperref}

\def \beq  {\begin{equation}}
\def \eeq  {\end{equation}}
\def \ber  {\begin{eqnarray}}
\def \eer  {\end{eqnarray}}

\begin{document}
\newcommand{\newc}{\newcommand}

\newc{\be}{\begin{equation}}
\newc{\ee}{\end{equation}}
\newc{\ba}{\begin{eqnarray}}
\newc{\ea}{\end{eqnarray}}
\newc{\bea}{\begin{eqnarray*}}
\newc{\eea}{\end{eqnarray*}}
\newc{\D}{\partial}
\newc{\ie}{{\it i.e.} }
\newc{\eg}{{\it e.g.} }
\newc{\etc}{{\it etc.} } 
\newc{\etal}{{\it et al.}}
\newc{\lcdm}{$\Lambda$CDM }
\newcommand{\nn}{\nonumber}
\newc{\ra}{\Rightarrow}

\title{Primordial Power Spectra of Cosmological Fluctuations with Generalized Uncertainty Principle and Maximum Length Quantum Mechanics} 
\author{F. Skara}\email{fskara@cc.uoi.gr}
\author{L. Perivolaropoulos}\email{leandros@uoi.gr} 
\affiliation{Department of Physics, University of Ioannina, 45110 Ioannina, Greece}

\date {\today} 

\begin{abstract}
The existence of the cosmological particle horizon as the maximum measurable length $l_{max}$ in the universe leads to a generalization of the quantum uncertainty principle (GUP) to the form $\Delta x \Delta p \geq \frac{\hbar}{2}\frac{1}{1-\alpha\Delta x^2} $, where $\alpha\equiv l_{max}^{-2}$. The implication of this GUP and the corresponding generalized commutation relation  $[x,p] =i\hbar \frac{1}{1-\alpha x^2}$ on simple quantum mechanical systems has been discussed recently \cite{Perivolaropoulos:2017rgq} by one of the authors and shown to have extremely small (beyond current measurements) effects of the energy spectra of these systems due to the extremely large scale of the current particle horizon. This may not the case in the Early Universe during the quantum generation of the inflationary primordial fluctuation spectrum. Here we estimate the effects of such GUP on the primordial fluctuation spectrum and on the corresponding spectral index. In particular motivated by the above GUP we generalize the field commutation (GFC) relation to $[\varphi(\bold{k}),\pi_{\varphi}(\bold{k'})]=i\delta(\bold{k}-\bold{k'})\frac{1}{1-\mu\varphi^2(\bold{k})}$, where $\mu\simeq\alpha^2\equiv l_{max}^{-4}$ is a GFC parameter, $\varphi$ denotes a scalar field and $\pi_{\varphi}$ denotes its canonical conjugate momentum. In the context of this GFC we use standard methods to obtain the primordial scalar perturbations spectrum and show that it is of the form $P_S(k)=P_S^{(0)}(k)\left(1+\frac{\bar{\mu}}{k}\right)$ where $\bar{\mu}\equiv\mu V_* \simeq \sqrt{\alpha}= l_{max}^{-1}$ (here $V_*\simeq l_{max}^3$ is the volume corresponding to  the maximum measurable scale $l_{max}$) and $P_S^{(0)}(k)$ is the standard primordial spectrum obtained in the context of the Heisenberg uncertainty principle (HUP $\mu=0$). We show that the scalar spectral index predicted by the model,  defined from $P_S(k)=A_Sk^{n_s -1}$ is running and may be written as  $n_s=1-\lambda-\frac{\bar{\mu}}{k}$ with $\lambda=6\epsilon-2\eta$ (where $\epsilon$ and $\eta$ are the slow-roll parameters). Using observational constraints on the scale dependence of the spectral index $n_s$ a cosmological constraint may be imposed on $\bar{\mu}$ as $\bar{\mu}=(0.9\pm 7.6)\cdot 10^{-6} h/Mpc$. Using this result we estimate the GUP parameter $\alpha\lesssim 10^{-54} m^{-2}$ at $1\sigma$ and $\alpha\lesssim 10^{-52} m^{-2}$ at $2\sigma$. The $2\sigma$ range od $\alpha$  corresponds to $l_{max}\gtrsim 10^{26} m $ which is of the same order as the current particle horizon. Thus the assumption that a maximum measurable length could emerge as a result of presence of the cosmological particle horizon remains a viable assumption at the $2\sigma$ level. 
\end{abstract} 
\maketitle
  
\section{Introduction}  
\label{sec:Introduction} 
 
A central issue of fundamental research is the  unification of quantum theory (QT) and general relativity (GR) in the framework of quantum  gravity (QG). A critical scale in the context of this unification is the Planck scale defined as $l_{pl}=\sqrt{\frac{\hbar G}{c^3}}=10^{-35}m$ (see Ref.\cite{Garay:1994en} for a review) which has been shown to be the minimum measurable scale if both QT and GR are applicable. Indeed it may be shown \cite{Plato:2016azz} that the high energies required to probe scales smaller than the Planck scale would lead to the formation of a black hole through the gravitational disturbances of spacetime structure which would prohibit any measurement on smaller scales. The existence of such a minimum measurable length would lead to a modification of the Heisenberg Uncertainty Principle \cite{aHeisenberg:1927zz,Robertson:1929zz}(HUP) to the so-called Generalized (Gravitational) Uncertainty Principle (GUP)(see Ref.\cite{Tawfik:2015rva} for a review)
\be
\Delta x \Delta p \geq \frac{\hbar}{2} (1+\beta \Delta p^2)
\label{gup1}  
\ee 
where $\beta$ is  the GUP parameter defined as $\beta =\beta_0 /M_{pl} c^2=\beta_0 l_{pl}^2/\hbar^2$,  $M_{pl}c^2=10^{19}GeV$, $l_{pl}$ is the 4-dimensional fundamental Planck scale and $\beta_0$ is a dimensionless parameter expected to be of order unity.
Such a GUP is closely related to the concept of noncommutative geometry \cite{Connes:1994yd} and has been extensively investigated in Refs. \cite{Mead:1964zz,Maggiore:1993kv,Maggiore:1993rv,Maggiore:1993zu,Kempf:1994su,Hinrichsen:1995mf,Kempf:1996ss,Kempf:1996nm,Snyder:1946qz,Yang:1947ud,Karolyhazy:1966zz,Ashoorioon:2004vm,Ashoorioon:2004rs,Ashoorioon:2004wd,Ashoorioon:2005ep,Faizal:2014mba,Ali:2015ola,Mohammadi:2015upa,Faizal:2016zlo,Zhao:2017xjj}. In particular interest in a minimum measurable length or equivalently in a ultraviolet cutoff has been motivated by studies of  string theory \cite{Veneziano:1986zf,Gross:1987kza,Gross:1987ar,Amati:1987wq,Amati:1988tn,Konishi:1989wk,Kato:1990bd}, loop quantum gravity \cite{Rovelli:1989za,Rovelli:1994ge,Carr:2011pr,rovelli_2004,Ashtekar:2004eh,Thiemann:2006cf,Thiemann:2002nj}, quantum geometry \cite{Capozziello:1999wx}, doubly special relativity (DSR)\cite{AmelinoCamelia:2000mn,Cortes:2004qn,Magueijo:2001cr,Magueijo:2002am,AmelinoCamelia:2002wr,Magueijo:2004vv} and by  black hole physics \cite{Maggiore:1993rv,Adler:2001vs,Nozari:2011gj,Alasfar:2017loh} or even Gedanken experiments \cite{Scardigli:1999jh} and thermodynamic  properties  of  gravity \cite{Zhu:2008cg}. Several phenomenological implications of minimal length theories and quantum gravity phenomenology  were investigated and a number of researchers have studied phenomenological aspects of GUP effects  in several contexts (e.g. in Refs. \cite{Das:2008kaa,Das:2009hs} atomic physics experiments such as Lamb’s shift and  Landau levels have been considered and constraints on the minimum length scale parameter $\beta$ have been estimated ). In Refs. \cite{Ali:2009zq,Ali:2011fa,Nozari:2012gd,Das:2010zf,Basilakos:2010vs} a model that is consistent with string theory, black hole physics and DSR is presented and discussed. This model of GUP predicts both a minimal observable length and a maximal momentum simultaneously \cite{Ali:2011fa,Das:2011tq}.
  
The existence of a minimum measurable length is closely related to the existence of the black hole horizon which tends to form if length scales below the Planck scale are probed. Correspondingly, there is a maximum measurable length associated with the cosmological particle horizon \cite{Faraoni:2011hf,Davis:2003ze} which provides due to causality a maximum measurable length scale in the Universe. The particle horizon corresponds to the length scale of the boundary between the observable and the unobservable regions of the universe. This scale at any time defines the size of the observable universe. The physical distance to this maximum observable scale at the cosmic time $t$ is given by (see e.g \cite{Kolb:1990vq,Hobson:2006se})
\be
l_{max}(t)=a(t)\int_0^t \frac{c\; dt}{a(t)}
\label{parthorscale}
\ee
where $a(t)$ is the cosmic scale factor.
For the best fit \lcdm cosmic background at the present time $t_0$ we have
\be
l_{max}(t_0)\simeq 14 Gpc \simeq 10^{26} m 
\label{parthort0}
\ee 

\begin{figure}
\begin{centering}
\includegraphics[width=0.45 \textwidth]{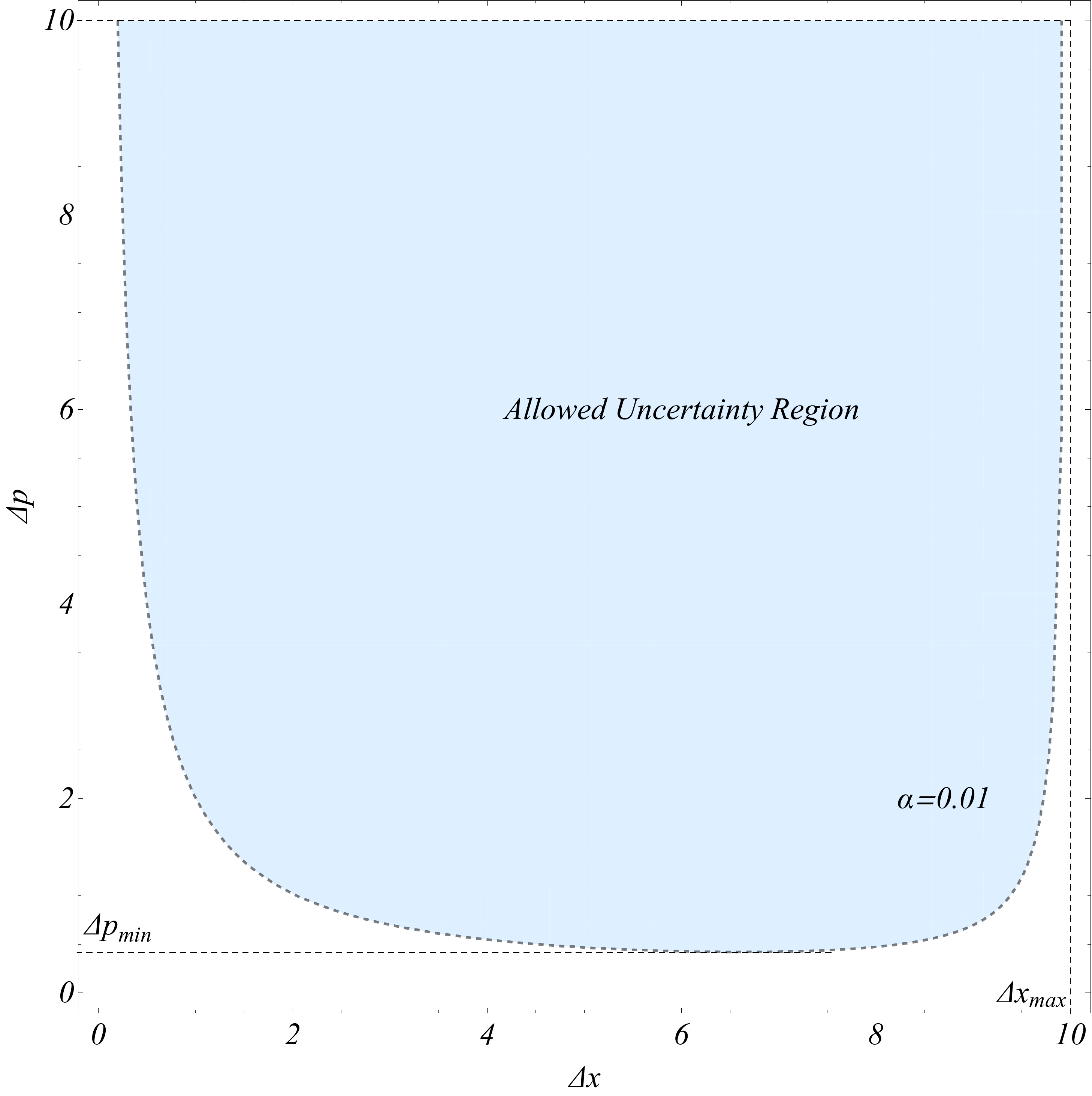}
\par\end{centering}
\caption{The deformation of the HUP in accordance with eq. (\ref{gupmaxlength}) after rescaling to dimensionless form using a characteristic length scale of the quantum system (from Ref. \cite{Perivolaropoulos:2017rgq}). } 
\label{lmax}
\end{figure}

This existence of such a maximum measurable length would lead to modified version of the GUP of the form \footnote{A perturbative version of this GUP  was introduced in \cite{Park:2007az} as  $\Delta x \Delta p \geq 1+ \alpha\frac{ \Delta x^2 }{L_*^2}$  (where $\alpha$ is a constant of order unity and $L_*$ is the characteristic, large length scale) and called extended uncertainty principle (EUP) by many authors \cite{Park:2007az,Bambi:2007ty,Zhu:2008cg,Mignemi:2009ji,Ghosh:2009hp,COSTAFILHO2016367,Schurmann:2018yuz,Mureika:2018gxl,Dabrowski:2019wjk}. Here we keep the notation `GUP' instead of `EUP' for consistency with Ref. \cite{Perivolaropoulos:2017rgq}.}  \cite{Perivolaropoulos:2017rgq} 
\be
\Delta x \Delta p \geq \frac{\hbar}{2}  \frac{1}{1- \alpha \Delta x^2}
\label{gupmaxlength}
\ee
As shown in Fig. \ref{lmax}, this GUP indicates the  existence of maximum position uncertainty (see Ref.\cite{Perivolaropoulos:2017rgq}) 
\be 
l_{max}\equiv \Delta x_{max} = \alpha^{-1/2}
\ee
due to the divergence of the RHS of eq. (\ref{gupmaxlength}). As shown in Fig. \ref{lmax} the existence of a maximum length scale is associated with the presence of a minimum momentum scale $\Delta p_{min}$. 
 
The GUP (\ref{gupmaxlength}) originates from a commutation relation of the form
\be
[x,p]=i \hbar \frac{1}{1-\alpha x^2}
\label{comrelgupmaxlength}
\ee  

It is straightforward to show (see in Appendix \ref{Appendix}) that this commutation relation leads to the GUP (\ref{gupmaxlength}) using the general uncertainty principle for the  pair of non-commuting observables $x$, $p$ 
\be 
\Delta x\Delta p \geq\frac{\hbar}{2}\mid\left\langle\left[\hat{x},\hat{p}\right]\right\rangle\mid
\label{genupAB}
\ee
with
\be
\Delta x\equiv \sqrt{\left\langle \left(\hat{x}-\left\langle \hat{x}\right\rangle\right)^2\right\rangle} 
\ee
\be
\Delta p\equiv \sqrt{\left\langle \left(\hat{p}-\left\langle \hat{p}\right\rangle\right)^2\right\rangle} 
\ee
where $\hat{x},\hat{p}$ are the operator representations of the observables $x$, $p$ .

The commutation relation (\ref{comrelgupmaxlength}) may be represented as shown in Appendix \ref{Appendix} by position and momentum operators of the form
\ba
p &=& \frac{1}{1 - \alpha x_0^2}p_0=(1+\alpha x_0^2 + \alpha^2 x_0^4 + ...)p_0
\label{reproperp} \\
x &=& x_0 
\label{reproperx} 
\ea
where $x_0$ and $p_0$ are the usual position and momentum operators satisfying the Heisenberg commutation relation $[x_0,p_0]=i\hbar$.

The representation (\ref{reproperp}), (\ref{reproperx}) may be used to solve the Schrodinger equation for simple quantum systems to find the dependence of the energy spectrum on the maximum measurable scale $l_{max}$.  Such an analysis has indicated \cite{Perivolaropoulos:2017rgq} that the current cosmic particle horizon is too large to lead to any observable effects in present day quantum systems. This however is not necessarily the case in the Early Universe when the particle horizon scale is much smaller and could leave an observable signature in the quantum generation of the primordial fluctuations during inflation. Thus, in the present analysis we wish to address the following questions
\begin{itemize}
\item
What is the deformation of the scale invariant spectrum of perturbations produced during inflation due to the Heisenberg algebra deformation (\ref{comrelgupmaxlength}) corresponding to the existence of a maximum measurable scale?
\item
What constraints can be imposed on the fundamental parameter $\alpha=l_{max}^{-2}$ from the observed power spectrum of primordial fluctuations?
\end{itemize}
The structure of this paper is the following: In the next section \ref{QHO}  we consider a simple harmonic oscillator in the presence of a large maximum measurable scale and find the variance of the position as a function of the parameter $\alpha$ and the corresponding variance in the context of the HUP ($\alpha=0$). In section \ref{PSCF} we generalize this analysis to the case of systems with infinite degrees of freedom (fields) and derive the spectrum and the spectral index of tensor and scalar perturbations generated during inflation as a function of the parameter $\alpha$ and of the corresponding spectrum obtained in the context of the HUP. In section \ref{OBCON} we use the derived theoretical expression for the (running) spectral index along with the corresponding observationally allowed range of the index as a function of the scale $k$ to derive constraints on the fundamental parameter $\alpha$ of the GUP. Finally in section \ref{Discussion} we conclude, summarize and discuss the implications and possible extensions of our analysis.\\

\section{Toy Model: The position variance of the Harmonic Oscillator under GUP}
\label{QHO}

In order to quantize the simple harmonic oscillator under the assumption of the GUP (\ref{gupmaxlength}) we need to generalize the expressions of the creation and annihilation operators $\hat{a}^{\dagger}$ and $\hat{a}$  in terms of $x,p$ so that the commutation relation \cite{Sakurai:1167961}
\be
[\hat{a},\hat{a}^{\dagger}]=1
\label{comrelaad}
\ee
is retained while at the same time the GUP commutation relation (\ref{comrelgupmaxlength}) is also respected. Thus, in order to satisfy these conditions, we generalize the analysis of Refs. \cite{Camacho:2003dm,Nozari:2005it} which applies to the  GUP (\ref{gup1}) and define 
\ba
\hat{a}&=&\frac{1}{\sqrt{2\hbar \omega}}\left(\omega\left[x+f(\alpha,x)\right]+ip\right)
\label{anngup} \\
\hat{a}^{\dagger}&=&\frac{1}{\sqrt{2\hbar \omega}}\left(\omega\left[x+f(\alpha,x)\right]-ip\right)
\label{cregup}
\ea
where $f(\alpha,x)$ is a function chosen so that the commutation relations (\ref{comrelaad}) and (\ref{comrelgupmaxlength}) are respected.

It is straightforward to show that the following function satisfies the aforementioned conditions simultaneously
\be
f(\alpha,x)= \sum_{n=1}^{\infty}\frac{(-\alpha)^n}{2n+1}x^{2n+1} 
\label{fx}
\ee
while it reduces to 0 in the limit $\alpha\rightarrow 0$ as it should.

Thus, we can rewrite eqs.(\ref{anngup}) and (\ref{cregup}) as
\be  
\hat{a}=\frac{1}{\sqrt{2\hbar \omega}}\left(\omega\frac{1}{\sqrt{\alpha}}arctan(\sqrt{\alpha}x)+ip\right)
\label{anntan}
\ee
\be
\hat{a}^{\dagger}=\frac{1}{\sqrt{2\hbar \omega}}\left(\omega\frac{1}{\sqrt{\alpha}}arctan(\sqrt{\alpha}x)-ip\right)
\label{cretan}
\ee
and the $p$ and $x$ operators are
\be 
p=-i\sqrt{\frac{\hbar \omega}{2}}\left(\hat{a}-\hat{a}^{\dagger}\right)
\ee
\be 
x=\frac{1}{\sqrt{\alpha}}tan \left(\sqrt{\frac{\hbar \alpha}{2\omega}}(\hat{a}+\hat{a}^{\dagger})\right)
\label{operxtan}
\ee\

Using  $tanx=x+\frac{x^3}{3}+\frac{2x^5}{15}+ ...$  , we have 
\be 
x=x_0+\frac{\alpha x_0^3}{3}+\frac{2\alpha^2 x_0^5}{15}+ ... 
\ee
where 
\be 
x_0=\sqrt{\frac{\hbar}{2\omega}}(\hat{a}+\hat{a}^{\dagger})
\ee
is the position operator in the case of the HUP ($\alpha=0$). 
Keeping the lower order terms in $\alpha$ (assuming $\frac{\alpha \hbar}{6\omega} \ll 1$) we obtain
\be 
x=x_0+\frac{\alpha x_0^3}{3}\Rightarrow x=\sqrt{\frac{\hbar}{2\omega}}(\hat{a}+\hat{a}^{\dagger})\left[1+\frac{\alpha \hbar}{6\omega}(\hat{a}+\hat{a}^{\dagger})^2 \right] 
\label{operxa}
\ee
For $\alpha=0$ we have 
\be 
x_0=\upsilon(\omega,t)\tilde{a}+\upsilon^*(\omega,t)\tilde{a}^{\dagger}
\label{operxo}
\ee
where 
\be 
\upsilon(\omega,t)=\sqrt{\frac{\hbar}{2\omega}}e^{-i\omega t}
\ee
is the properly normalized solution of the classical evolution equation of the harmonic oscillator $\frac{d^2\upsilon}{dt^2}+ \omega^2 \upsilon = 0 $.
Therefore the position operator may be expressed as
\be 
x=\left(\upsilon\tilde{a}+\upsilon^*\tilde{a}^{\dagger}\right)\left[1+\frac{\alpha}{3}(\upsilon\tilde{a}+\upsilon^*\tilde{a}^{\dagger})^2\right]
\ee
Thus the variance of the position in the ground state takes the form
\begin{widetext}
\be 
\langle |x|^2\rangle\equiv\langle 0| x^\dagger x |0\rangle\Rightarrow \langle |x|^2\rangle=|\upsilon(\omega,t)|^2\left[1+2\alpha|\upsilon(\omega,t)|^2\right]
\label{qfoperx}
\ee 
\end{widetext} 
which reduces to the familiar result for $\alpha=0$ (see e.g. \cite{Dodelson:2003ft,Baumann:2009ds}).

In the next section we generalize the above analysis to the case of quantum field fluctuations involving infinite degrees of freedom aiming to derive the perturbation power spectrum generated during inflation in the context of the GUP.\\

\section{PRIMORDIAL SPECTRA OF COSMOLOGICAL FLUCTUATIONS with GUP}  
\label{PSCF}

According to the decomposition theorem \cite{Lifshitz:1945du} the perturbations of each type evolve independently (at the linear level) and we can treat tensor (T) and scalar (S) perturbations of the metric separately. Therefore for spatially flat Friedmann-Robertson-Walker (FRW) background plus the perturbations we can write 
\be
ds_T^2=a^2\left[-d\tau^2+(\delta_{ij}+H_{ij})dx^idx^j\right]
\ee
and in conformal Newtonian gauge \cite{Mukhanov:1990me} 
\be 
ds_S^2=a^2\left[-(1+2\Psi)d\tau^2+\delta_{ij}(1+2\Phi)dx^idx^j\right]
\ee
where $a$ is the scale factor, $\tau$ is the conformal time, $\Psi$ corresponds to the gravitational potential of the perturbations, $\Phi$ is the perturbation of the spatial curvature\footnote{In the absence of anisotropic stress ($\Pi=0$) we have $\Psi=-\Phi$ \cite{Bertschinger:2001is}} and $H_{ij}$  is the tensor perturbation which has the form \footnote{It has this form in a coordinate system where wavevector $\bold{k}$ points along the z-axis.}
 \be 
\left[H_{ij}\right]=\left[
         \begin{array}{ccc} 
           h_+ & h_{\times} & 0  \\
           h_{\times} &-h_+ & 0 \\
           0 & 0 & 0 \\
         \end{array}
       \right]
\ee 
The classical evolution equations for the tensor mode perturbations $h_T$  (where $T={+},{\times}$ for two polarization states \cite{Misner:1974qy}) of the FRW metric during inflation in conformal time are obtained from the linearized Einstein equations and may be written as \cite{Grishchuk:1974ny} 
\be 
h_T''+ 2 \frac{a'}{a}h_T'+k^2h_T=0
\label{tfh}
\ee
where primes denote derivatives with respect to $\tau$. This becomes a simple harmonic oscillator equation by defining 
\be 
\tilde{h}_T\equiv\frac{a h_T}{\sqrt{16\pi G}}
\ee
and eq. (\ref{tfh}) takes the form
\be
\tilde{h}_T''+\omega^2\tilde{h}_T=0
\label{qfha}
\ee
where 
\be 
\omega^2 =k^2-\frac{a''}{a}
\ee
During slow roll inflation when the Hubble rate $H$ is nearly constant \cite{Lyth:1994dc}, the conformal time is \cite{Lyth:1998xn,Baumann:2009ds}
\be
\tau\simeq\frac{-1}{aH}
\label{ct}
\ee
Thus we obtain 
\be 
\omega^2 =k^2-\frac{2}{\tau^2}
\ee
We now quantize the tensor field fluctuations by promoting them to operators and imposing a generalized field commutation (GFC) relation \cite{Matsuo:2005fb,Kober:2011uj} corresponding to (\ref{comrelgupmaxlength}). This GFC takes the form ($\hbar=1$) 
\be
[\tilde{h}_T(\bold{k}),\pi_{\tilde{h}_T}(\bold{k'})]=i\delta(\bold{k}-\bold{k}')\frac{1}{1-\mu\tilde{h}_T^2(\bold{k})}
\label{guph} 
\ee
where $\pi_{\tilde{h}_T}$ is the conjugate momentum to $\tilde{h}_T$ which is given by
\be
\pi_{\tilde{h}_T}=\tilde{h}_T'-\frac{a'}{a}\tilde{h}_T 
\ee
and $\mu$ is a GFC parameter 
\be 
\mu \simeq \alpha^2 = l_{max}^{-4}
\label{param}
\ee
where $\alpha$ is the  parameter of the GUP (\ref{gupmaxlength}).
Thus we have an infinite number of decoupled harmonic oscillators corresponding to eq. (\ref{qfha}) which may be quantized in accordance with the GFC (\ref{guph}).
Using the results of the previous section we connect the field normal modes with the creation and annihilation operators which satisfy the commutation relation $[\hat{a}_{\bold{k}},\hat{a}_{\bold{k'}}^{\dagger}]=\delta^3(\bold{k}-\bold{k'})$, as  
\be 
\tilde{h}_T(\bold{k})=\frac{1}{\sqrt{\mu}}tan \left(\sqrt{\frac{\mu}{2\omega}}(\hat{a}_{\bold{k}}+\hat{a}_{\bold{k}}^{\dagger})\right)
\ee
\be 
\pi_{\tilde{h}_T}(\bold{k})=-i\sqrt{\frac{\omega}{2}}\left(\hat{a}_{\bold{k}}-\hat{a}_{\bold{k}}^{\dagger}\right)
\ee
and obtain the variance of the perturbations as
\begin{widetext}   
\be 
\langle h_T^\dagger(\bold{k},\tau)h_T(\bold{k'},\tau)\rangle = \frac{16\pi G}{a^2}|\upsilon(\bold{k},\tau)|^2\left[1+2\bar{\mu}|\upsilon(\bold{k},\tau)|^2\right](2\pi)^3 \delta^3(\bold{k}-\bold{k'})\equiv (2\pi)^3 P_h(k)\delta^3(\bold{k}-\bold{k'})
\label{psh}
\ee
\end{widetext}
where $P_h$ is the power spectrum of the primordial tensor perturbations of the metric, the Dirac delta function enforces the independence of the different modes ($ h(\bold{k},\tau)$ is uncorrelated with $h(\bold{k'},\tau)$ if  $\bold{k}\neq\bold{k}'$ ) and
\be 
\bar{\mu}=\mu V_*
\label{barmu}
\ee
Here the volume scale $V_*=\delta^3(0)\simeq l_{max}^3$ is an infrared regulator \cite{Oblak:2016eij} while $\upsilon$ satisfies the Mukhanov-Sasaki equation \cite{Mukhanov:1988jd,Kodama:1985bj,Stewart:1993bc}
\be 
\upsilon''(k,\tau)+(k^2-\frac{a''}{a})\upsilon (k,\tau)=0
\label{ups}
\ee
During slow-roll inflation with initial condition $\upsilon (k,\tau)=\frac{1}{\sqrt{2k}}e^{-ik\tau}$ and by virtue of eq. (\ref{ct}) (as in spatially flat de Sitter background) we obtain the Bunch-Davies  solution of eq. (\ref{ups})  \cite{Birrell:1982ix,Parker:2009uva,Kinney:2009vz,Baumann:2009ds}
\be 
\upsilon(k,\tau)=\frac{e^{-ik\tau}}{\sqrt{2k}}\left(1-\frac{i}{k\tau}\right)
\label{mod}
\ee
Using eq. (\ref{psh}) we can write the primordial power spectrum for tensor modes as 
\be 
P_h(k)=P_h^{(0)}(k)\left[1+\frac{\bar{\mu}a^2}{8\pi G} P_h^{(0)}(k)\right]
\label{psh0}
\ee 
where 
\be 
P_h^{(0)}(k)= \frac{16\pi G}{a^2}|\upsilon(k,\tau)|^2
\label{psh00}
\ee
Once  $k|\tau|<1 $ , the mode leaves the horizon, after which $h$ remains constant. Thus, using eqs. (\ref{mod}) and (\ref{psh00}) we obtain
\be 
P_h^{(0)}(k)= \frac{16\pi G}{a^2}\frac{1}{2k^3\tau^2}=\frac{8\pi G H^2}{k^3}
\ee
where the equality on the second line holds because we have assumed that $H$ is constant and $\tau=-\frac{1}{a H}$.\footnote{We evaluate $H$ at the time when the mode leaves the horizon.}

In a similar manner we may investigate scalar perturbations induced by quantum fluctuations of the inflaton scalar field  $\phi$ \cite{Liddle:1993fq,Lidsey:1995np,Baumann:2009ds} of the form 
\be 
\phi(\bold{x},t)=\phi^{(0)}(t)+\delta\phi(\bold{x},t)
\ee 
where  $\phi^{(0)}$ is the zero-order part and  $\delta\phi$ is the first-order perturbation.

The fluctuations $\delta\phi$ of the scalar field driving inflation evolve in conformal time $\tau$  according to the equation (see e.g. \cite{Kolb:1990vq})
\be 
\delta\phi''+ 2 \frac{a'}{a}\delta\phi'+k^2\delta\phi=0
\label{ifphi}
\ee
Using the definition
\be 
\varphi=a\delta\phi
\ee
eq. (\ref{ifphi}) becomes
\be
\varphi''+\omega^2\varphi=0
\label{qfha2}
\ee
with $\omega^2 =k^2-\frac{a''}{a}$.

In the context of the maximal measurable length GUP as applied to the case of the inflaton fluctuations,  the field commutation relation gets generalized as 
\be
[\varphi(\bold{k}),\pi_{\varphi}(\bold{k'})]=i\delta(\bold{k}-\bold{k'})\frac{1}{1-\mu\varphi^2(\bold{k})}
\label{gupphi} 
\ee
where $\pi_{\varphi}$ is the conjugate momentum to $\varphi$ which is given by 
\be
\pi_{\varphi}=\varphi'-\frac{a'}{a}\varphi
\ee  

Since eq. (\ref{ifphi}) is identical to eq. (\ref{tfh}) we can use the result of eq. (\ref{psh0}) without the factor $16\pi G$ in order to turn the dimensionless $h$ into a field $\delta\phi$ with dimensions of mass
\be 
P_{\delta\phi}(k)=P_{\delta\phi}^{(0)}(k)\left[1+2\bar{\mu}a^2 P_{\delta\phi}^{(0)}(k)\right] 
\label{psphi0}
\ee
where  
\be 
P_{\delta\phi}^{(0)}(k)= \frac{H^2}{2 k^3} 
\ee
In the case $\bar{\mu}=0$ eqs. (\ref{psh0}) and (\ref{psphi0}) reduce to the familiar results of HUP \cite{Mukhanov:1990me}.

The perturbation from the scalar field driving inflation $\delta\phi$ gets transferred to the gravitational potential $\Phi$. The post inflation power spectrum of $\Phi$ is related to the horizon-crossing power spectrum of $\delta\phi$ via  \cite{Dodelson:2003ft}
\be
P_\Phi=\frac{16\pi G}{9\epsilon}P_{\delta\phi}
\ee 
where $\epsilon$ is the Hubble slow-roll parameter, defined as  
\be 
\epsilon\equiv\frac{d}{dt}\left(\frac{1}{H}\right)
\label{epsilon}
\ee
We note that the Hubble slow-roll parameter $\epsilon$ is equal to the first potential slow-roll parameter $\epsilon_V$, to leading order in the slow-roll approximation \cite{Liddle:2000cg,Liddle:1994dx,Liddle:1992wi,Lyth:1998xn,Baumann:2009ds}
\be
\epsilon\simeq\epsilon_V\equiv \frac{1}{16\pi G}(\frac{V'}{V})^2
\label{epsilonV}
\ee
where $V'$ is defined as the derivative of the potential $V$ with respect to the field $\phi^{(0)}$.

In the case of single-field  slow-roll models of inflation for modes which are outside the horizon ($k|\tau|\ll 1$) at the end of inflation, the primordial spectra of scalar and tensor perturbations do not depend on time\footnote{We assume that non-adiabatic pressure terms are negligible.} and it is conventional to write \cite{Lyth:1998xn}

\be 
P_S(k)\equiv k^3 P_\Phi(k)\equiv A_Sk^{n_s -1}
\label{psr}
\ee
\be 
P_T(k)\equiv k^3 P_{h}(k)\equiv A_Tk^{n_T}
\label{pst}
\ee
where $A_S (A_T)$ is the scalar (tensor) amplitude and $n_s (n_T)$ is the scalar (tensor) spectral index. The special case with $n_s= 1$ ($n_T=0$) results in the scale-invariant spectrum. 

From eqs. (\ref{psh0}) and (\ref{pst}) we obtain
\be 
P_T(k)=P_T^{(0)}(k)\left[1+\frac{\bar{\mu} a^2}{8\pi G k^3} P_T^{(0)}(k)\right]
\label{pst0}
\ee 
where (for $k|\tau|\ll 1$ )  
\be 
P_T^{(0)}(k)= \frac{8\pi G}{a^2\tau^2}=8\pi G H^2
\label{ptkmo}
\ee
It is straightforward to show  at the horizon crossing time ($k=a H$) 
\be
P_T(k)=P_T^{(0)}(k)\left(1+\frac{\bar{\mu}}{k}\right)
\label{ptkm}
\ee
In eq. (\ref{pst}) the tensor spectral index is defined  as
\be
n_T\equiv\frac{d \ln P_T}{d \ln k} 
\label{lnnT}
\ee
Also by virtue of eq. (\ref{epsilon}) we have that the logarithmic derivative of Hubble rate at horizon crossing is 
\be 
\frac{d\ln H}{d\ln k}=-\epsilon
\ee
Therefore using eqs. (\ref{ptkmo}), (\ref{ptkm}) and (\ref{lnnT}) we obtain that the tensor spectral index runs as
\be
n_T=-2\epsilon-\frac{\bar{\mu}}{k}
\ee
Similarly, from eq. (\ref{psphi0}) and using  $P_S=k^3\frac{16\pi G}{9\epsilon}P_{\delta\phi}$ we obtain at horizon crossing time ($k=a H$)
\be 
P_S(k)=P_S^{(0)}(k)\left[1+\frac{9\bar{\mu} \epsilon}{8\pi G H^2 k} P_S^{(0)}(k)\right]
\label{pss0}
\ee
where   
\be 
P_S^{(0)}(k)= \frac{8\pi G H^2}{9\epsilon}
\label{pseo} 
\ee
It is straightforward to show that the     
\be 
P_S(k)=P_S^{(0)}(k)\left(1+\frac{\bar{\mu}}{k}\right) 
\label{psmo}
\ee 
Notice that Eqs. (\ref{pss0}) and (\ref{pseo}) have a generic form which could have been guessed even on the basis of dimensional analysis. However, here we have demonstrated in detail that these equations are not simply well motivated parametrizations based on dimensional analysis. Instead they constitute the unique and generic prediction of the inflationary power spectrum of  fluctuations generated in the context of the GUP eq. (\ref{gupphi}) as derived in the context of our analysis. Thus there is no room to modify eq. (\ref{pss0}) without violating the physical principle corresponding to the GUP (\ref{gupphi}).

In eq. (\ref{psr}) the scalar spectal index is defined as
\be 
n_s-1\equiv\frac{d \ln P_{\Phi}}{d \ln k} 
\label{lnnS}
\ee
Now using the eq. (\ref{epsilonV})  and  the Hubble slow-roll parameter \cite{Liddle:1992wi}
\be 
\delta\equiv\frac{1}{H}\frac{d^2\phi^{(0)}/dt^2}{d\phi^{(0)}/dt}
\ee
we have that the logarithmic derivative of the slow-roll parameter $\epsilon$ is 
\be 
\frac{d\ln \epsilon}{d\ln k}=2(\epsilon+\delta)
\ee
Therefore using eqs. (\ref{pseo}), (\ref{psmo}) and (\ref{lnnS}) we obtain that the scalar spectral index runs as
\be
n_s=1-4\epsilon-2\delta-\frac{\bar{\mu}}{k} 
\ee 

Alternatively using the the second potential slow-roll parameter  $\eta\equiv\frac{1}{8\pi G}\frac{V''}{V}$ and the relation  $\delta=\epsilon-\eta$\footnote{The second slow-roll parameter $\delta$ and the second potential slow-roll parameter $\eta$ are  sometimes defined as $\eta$ and $\eta_V$ respectively, so that the relation has the form $\eta=\epsilon_V-\eta_V$} \cite{Lyth:1998xn}, we obtain

\be
n_s=1-6\epsilon+2\eta-\frac{\bar{\mu}}{k}
\label{nsehm}   
\ee
In the next subsection we use  observational scalar spectral index data to obtain bounds on $\bar{\mu}$.

\begin{figure*}
\begin{centering}
\includegraphics[width=0.75\textwidth]{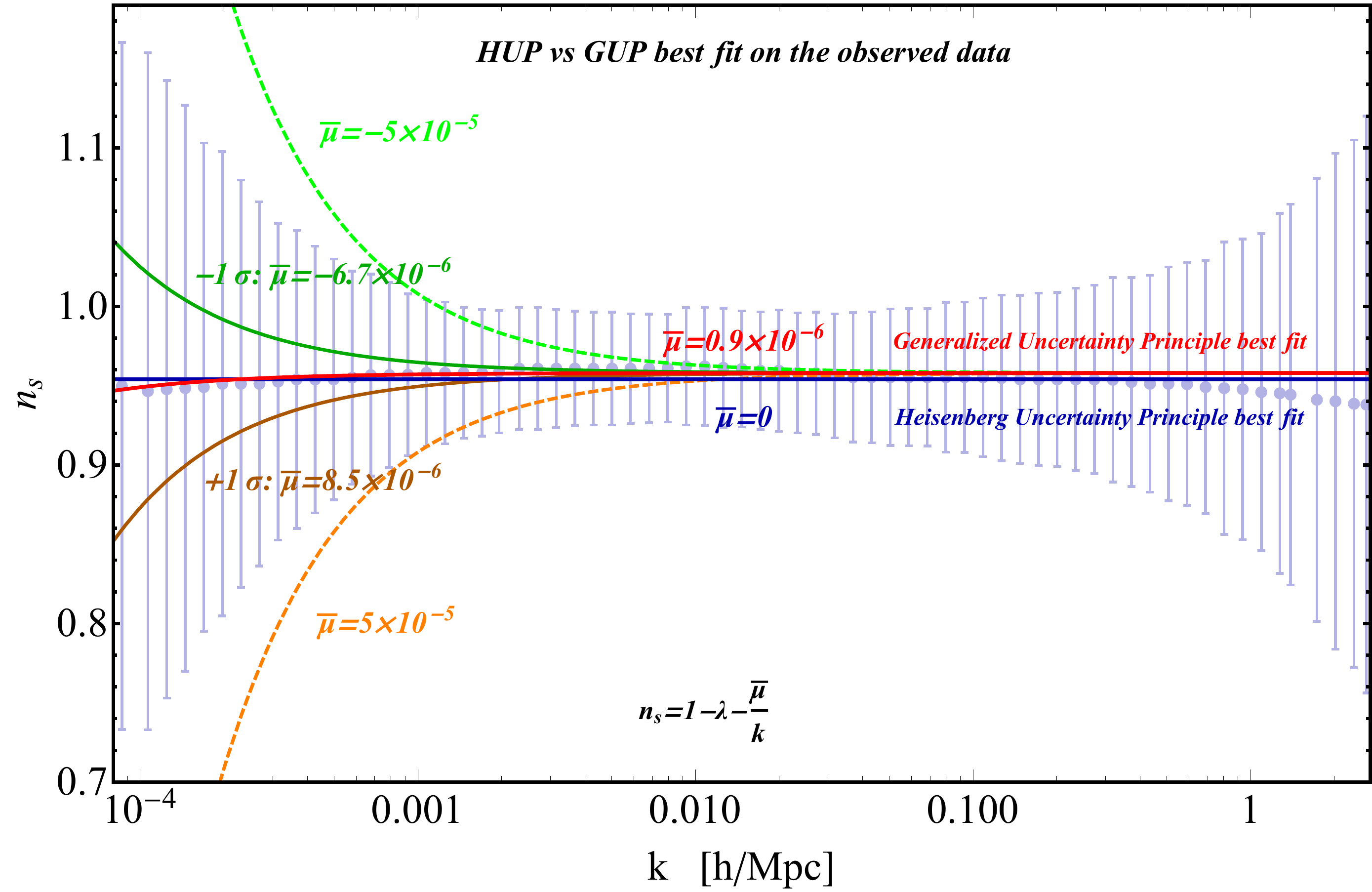}
\par\end{centering}
\caption{The best fit forms of the scalar spectral index eq. (\ref{nslm}) (blue curve for HUP and red curve for GFC eq. (\ref{gupphi})) on the observed data (thick dots). The green and brown continuous curves correspond to  $-1\sigma$ and $+1\sigma$ deviation of the parameter $\bar{\mu}$ respectively. The light green and the orange dashed curves correspond to observationally allowed range for the spectral index $n_S$ at approximatelly $2\sigma$ level. } \label{figerbf}
\end{figure*}

\section{Observational Constraints}
\label{OBCON}

The predicted form of the running spectral index eq. (\ref{nsehm}) reduces to the standard form \cite{Lyth:1994dc,Lyth:1998xn} for the HUP ($\bar{\mu}=0$) and may be used along with observational constraints of the spectral index to impose constraints on the GFC parameter $\bar{\mu}$.

The parameters that can lead to deviations from scale invariance of the spectral index are the GFC parameter $\mu$ and the slow-roll parameter $\lambda$ defined as

\be  
\lambda=6\epsilon-2\eta 
\ee

Thus using eq. (\ref{nsehm}), the scalar spectral index takes the form
 \be
n_s=1-\lambda-\frac{\bar{\mu}}{k} 
\label{nslm}
\ee

In order to impose constraints on the parameters $\lambda, \bar{\mu}$ we use constraints on the scalar spectral index of Ref. \cite{Peiris:2009wp} which are based on the angular power spectrum data of the 5 year Wilkinson Microwave Anisotropy Probe (WMAP5) Cosmic Microwave Background (CMB) temperature and polarization, the Large Scale Structure (LSS) data of the Sloan Digital Sky Survey (SDSS) data release 7 (DR7) Luminous Red Galaxy (LRG) power spectrum, and  the Lyman-alpha forest (Lya) power spectrum constraints. The allowed range on $n_s$ is shown in Fig. \ref{figerbf}.

Expressing this range as a set of $N=60$ datapoints leads to constraints on the parameters $\lambda, \bar{\mu}$ through the maximum likelihood method \cite{DBLP:books/daglib/0072312}. As a first step for the construction of $\chi^2$, we consider the vector \cite{ Verde:2009tu} 

\be
V^i(k_i,\lambda,\bar{\mu} )\equiv  n_{s,i}^{obs}(k_i)-n_{s,i}^{th}(k_i,\lambda,\bar{\mu})
\ee\
where $n_{s,i}^{obs}(k_i)$ and $n_{s,i}^{th}(k_i,\lambda,\bar{\mu})$ are the observational and the theoretical spectral index at wavenumber $k_i$ respectively ( $i=1,...,N$  with  $N$ corresponds to the number of datapoints).
Then we obtain $\chi^2$ as 
\be 
\chi^2=V^i F_{ij}V^j
\label{x2}
\ee
where $F_{ij}$ is the Fisher matrix \cite{10.2307/2342435} (the inverse of the covariance matrix $C_{ij}$ of the data).

The $N\times N$ covariance matrix  is assumed to be of the form
\be
\left[C_{ij}\right]=\left[
         \begin{array}{cccc}
           \sigma_1^2 & 0 & 0 & \cdots \\
           0 & \sigma_2^2 & 0& \cdots \\
           0 & 0 & \cdots &   \sigma_N^2 \\
         \end{array}
       \right] \label{eq:totalcijother}
\ee
where $\sigma_i$ denotes the $1\sigma$ error of data point $i$.

\begin{widetext} 
\begin{figure*}
\includegraphics[width=0.98\textwidth]{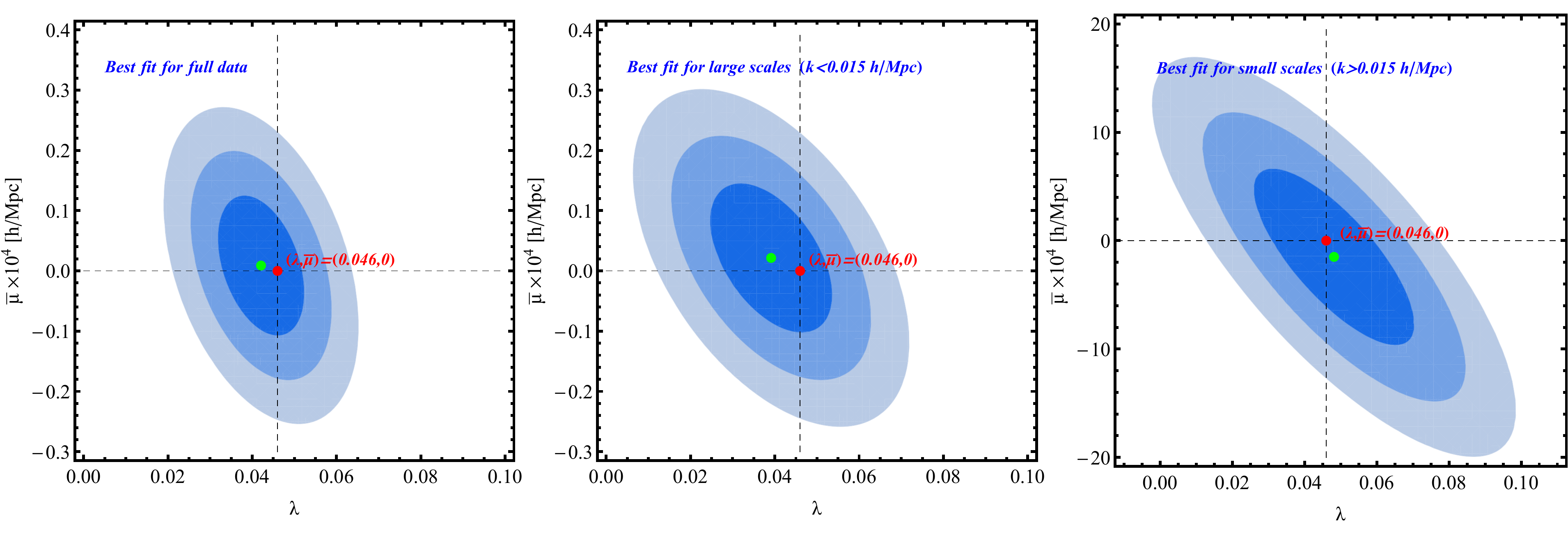}
\caption{The $1\sigma-3\sigma$ contours in the ($\lambda,\bar{\mu}$) parametric space. The contours describe the corresponding confidence regions obtained from the full data set (left panel), large scales ($k<0.015$   $h/Mpc$) data (middle panel), and small scales ($k>0.015$    $h/Mpc$) data (right panel). The red and green points correspond to the HUP and GUP best fits respectively.} \label{cont}
\end{figure*}
\end{widetext}

The $68.3\%$ ($1\sigma$), $95.4\%$ ($2\sigma$) and $99.7\%$ ($3\sigma$) confidence contours in the $\lambda$ and  $\bar{\mu}$  parametric space are shown  in  Fig. \ref{cont}. The contours correspond to confidence regions obtained from the full data set (left panel), the large scales ($k<0.015$  $h/Mpc$) data (middle panel), and the small scales ($k>0.015$   $h/Mpc$) data (right panel). The $1\sigma$-$3\sigma$ contours for parameters $\lambda$ and  $\bar{\mu}$ correspond to the curves $\chi^2(\lambda,\bar{\mu})=\chi^2_{min}+2.3$, $\chi^2(\lambda,\bar{\mu})=\chi^2_{min}+6.17$ and $\chi^2(\lambda,\bar{\mu})=\chi^2_{min}+9.21$ respectively. Notice  (in Fig. \ref{cont}) that the large scales are most efficient in constraining the GFC parameter $\bar{\mu}$. 
The largest scales that correspond to small $k$ give the largest value for the correction ${\bar \mu} /k$ of the power spectrum and the spectral index  eq. (\ref{nslm}).  Thus it is these scales that are more sensitive to the correction and lead to the strongest constraints as shown in Fig. \ref{cont}. 
 
In Table \ref{table:surveys} we show the best fit values of parameters $\lambda$ and  $\bar{\mu}$  with the corresponding $1\sigma$ standard deviations. In the case of HUP ($\bar{\mu}=0$) the result agrees with the current best fit values of the scalar spectral index from Planck which indicate that $\lambda\simeq 0.04$ \cite{Aghanim:2018eyx}  .

\begin{center}  
\begin{table}   
\centering  
\begin{tabular}{c c c c c c}
\hhline{======}
\multicolumn{6}{c}{ }\\
\multicolumn{6}{c}{GFC}\\
\multicolumn{6}{c}{ }\\
\hline 
   &&&&& \\
  Parameter &Full & &Large Scales &&Small Scales    \\
  & Data ($1\sigma$)&&Data ($1\sigma$) &&Data ($1\sigma$)\\
   &&&&& \\
 \hline 
    &&& \\
  $\bar{\mu} $ &$0.9 \pm 7.6$&  &$2.1 \pm 8.1$ & &$-149 \pm 535$   \\
   &$[\times 10^{-6}h/Mpc]$&&$[\times 10^{-6}h/Mpc]$& &$[\times 10^{-6}h/Mpc]$\\
    &&& \\
   $\lambda $ & $0.042\pm 0.0067$&&$0.039\pm 0.0095$&& $0.048\pm 0.0146$\\
 \hhline{======}   
\end{tabular} 
\caption{\small  The best fit values of parameters $\lambda$ and  $\bar{\mu}$ with the corresponding $1\sigma$ standard deviations for the fitted spectral index on the observed data \cite{Peiris:2009wp}.}
\label{table:surveys} 
\end{table} 
\end{center}
Using eq. (\ref{param}) and the $1\sigma$ constraint on the GFC parameter $\bar{\mu}\lesssim  10^{-5}h/Mpc$ we can obtain the single GUP free parameter  as
\be
\alpha=\bar{\mu}^2\lesssim 10^{-54}m^{-2}
\ee
and the corresponding maximum measurable scale as
\be
l_{max}=\bar{\mu}^{-1}\gtrsim 10^{27}m
\ee
This result is one order of magnitude larger than  the present day particle horizon ($l_{max}(t_0)\simeq 10^{26}m$) given in eq. (\ref{parthort0}). 
However, at about $2\sigma$ level the physically anticipated maximum measurable scale (the particle horizon scale) is included in the observationally allowed range of the maximum measurable scales. Thus, the emergence of the parameter $\mu$ in  (\ref{guph}) and (\ref{gupphi}) as a consequence of a maximum measurable length associated with the cosmological particle horizon remains an observationally viable hypothesis.  The parameter $\bar{\mu}$ is a fundamental parameter connected to the GUP (\ref{gupphi}) and it is not necessarily connected with the detailed physics of inflation.
Thus our analysis can not directly impose constraints on models of inflation even though there may be an indirect connection of the present day value of $l_{max}$ with the scale of inflation. Such a connection would require a time dependent form fo $l_{max}$ and is beyond the scope of the present analysis.

\section{CONCLUSIONS-DISCUSSION} 
\label{Discussion}

We have derived the generalized form of the primordial power spectrum of cosmological perturbations generated during inflation due to the quantum fluctuations of scalar and tensor degrees of freedom in the context of a generalization of quantum mechanics involving a maximum measurable length scale. The existence of such a scale is motivated by the existence of the particle horizon in cosmology and would lead to a generalization of the uncertainty principle (GUP) to the form  $\Delta x \Delta p \geq \frac{\hbar}{2}\frac{1}{1-\alpha\Delta x^2} $, which implies the existence of a maximum position and a minimum momentum uncertainty (infrared cutoff)\cite{Perivolaropoulos:2017rgq}.  The GUP implies a generalization of the commutation relation between conjugate operators including fields and their conjugate momenta. For example we showed that the generalized field commutation (GFC) relation between a scalar field and its conjugate momentum  $[\varphi(\bold{k}),\pi_{\varphi}(\bold{k}')]=i\delta(\bold{k}-\bold{k}')\frac{1}{1-\mu\varphi^2(\bold{k})}$ which is implied by the GUP leads to a modified primordial spectrum of scalar  perturbation are $P_S(k)=P_S^{(0)}(k)\left(1+\frac{\bar{\mu}}{k}\right)$ with a running spectral index of the form $n_s=1-\lambda-\frac{\bar{\mu}}{k}$ with $\lambda=6\epsilon-2\eta$.  

Using cosmological constraints of the scalar perturbations spectral index as a function of the scale $k$  \cite{Verde:2009tu}  we imposed constraints on the parameter of the GFC $\bar{\mu}\simeq l_{max}^{-1}$.  We found that $\bar{\mu}=(0.9\pm 7.6)\cdot 10^{-6} h/Mpc$ at the $1\sigma$ level which corresponds to an upper bound scale $l_{max}$ larger than the present horizon scale. At $2\sigma$ level we find that the observationally allowed range of $l_{max}$ includes the current cosmological horizon scale  $l_{max}\simeq 10^{26} m $. Thus at $2\sigma$ level, the derived observational constraints on $l_{max}$ are consistent with the physically anticipated maximum measurable scale which is the current cosmological particle horizon and are much more powerful than the corresponding constraints obtained using laboratory data measuring the energy spectrum of simple quantum systems obtained in Ref. \cite{Perivolaropoulos:2017rgq}.  

An interesting extension of our analysis would be the consideration of other types of GUP (e.g. the UV cutoff GUP of eq. (\ref{gup1})) and the derivation of constraints on the corresponding fundamental parameters using cosmological data and constraints on the power spectrum index. 

An alternative approach in deriving the effects of a  GUP on the primordial perturbation spectrum involves the generalization of the position and momentum operators as described in the Introduction, but with an ultraviolet rather than infrared cutoff, while keeping the field theoretical commutation relations unchanged \cite{Kempf:2000ac,Palma:2008tx}. According to \cite{Kempf:2000ac,Palma:2008tx}, this approach would also lead to a modification of the evolution of the field perturbation modes eq. (\ref{ifphi}) even though this equation is derived before quantization at the classical level. This approach is questionable as it is implemented at the classical level. Nevertheless, it would be of interest to extend our analysis to include such effects of modification of the classical evolution of field perturbations due to a generalization of position and momentum operators. \\

\textbf{Supplemental Material:} The Mathematica file used for the numerical analysis and for construction of the figures can be found in \cite{suppl}.\\

\section*{ACKNOWLEDGEMENTS}
This  article has benefited from COST Action CA15117 (CANTATA), supported by COST (European Cooperation in Science and Technology).\\

\appendix
 
\section{From generalized commutator to generalized uncertainty}
\label{Appendix}
We assume the commutation relation of the form
\be
[x,p]=i \hbar \frac{1}{1-\alpha x^2}\simeq i \hbar (1+\alpha x^2)
\label{comrelgupmaxlengthap}
\ee 
where the last approximate equality is applicable under the condition $\alpha x^2\ll 1 $. The commutation relation (\ref{comrelgupmaxlengthap}) may be represented by position and momentum operators of the form
\ba
p &=& \frac{1}{1 - \alpha x_0^2}p_0=(1+\alpha x_0^2 + \alpha^2 x_0^4 + ...)p_0
\label{reproperpap} \\
x &=& x_0 
\label{reproperxap} 
\ea
where $x_0$ and $p_0$ are the usual position and momentum operators satisfying the Heisenberg commutation relation $[x_0,p_0]=i\hbar$.
\begin{widetext}
The proof that the commutation relation (\ref{comrelgupmaxlengthap}) may be represented by position and momentum operators of the form (\ref{reproperpap}) and (\ref{reproperxap}) is
\be
[x,p]=[x_0,(1+\alpha x_0^2 + \alpha^2 x_0^4 + ...)p_0]= [x_0,p_0]+ \alpha x_0^2 [x_0,p_0]+\alpha^2 x_0^4 [x_0,p_0]+... =[x_0,p_0] \frac{1}{1- \alpha x_0^2}=i \hbar \frac{1}{1-\alpha x^2}
\ee
Also, the proof that the commutation relation (\ref{comrelgupmaxlengthap}) leads to a GUP of the form (\ref{gupmaxlength}) is
\be
\begin{matrix}
\Delta x \Delta p \geq \frac{\hbar }{2}<\frac{1}{1-\alpha x^2}> =\frac{\hbar }{2}(1+\alpha <x^2> + \alpha^2 <x^4> + ...)\geq \frac{\hbar }{2}(1+\alpha <x^2> + \alpha^2 <x^2>^2 + ...)=\\ 
\\
\frac{\hbar }{2}(1+\alpha({\Delta x}^2 + <x>^2) + \alpha^2 ({\Delta x}^2 + <x>^2)^2 + ...)=\frac{\hbar}{2} \frac{1}{1- \alpha ({\Delta x}^2+<x>^2)}\geq \frac{\hbar}{2} \frac{1}{1- \alpha {\Delta x}^2}\Rightarrow\\
\\
\Delta x \Delta p \geq  \frac{\hbar}{2} \frac{1}{1- \alpha {\Delta x}^2}\\
\end{matrix}
\ee
\end{widetext}

\bibliography{bibliography}              
 
\end{document}